# A hierarchical neural stimulation model for pain relief by variation of coil design parameters

**Raunaq Pradhan[1] and Yuanjin Zheng[1]**

[1] School of Electrical and Electronic Engineering, Nanyang Technological University, Singapore, Singapore-639798

**Abstract-** *Neural stimulation represents a powerful technique for neural disorder treatment. This paper deals with optimization of coil design parameters to be used during stimulation for modulation of neuronal firing to achieve pain relief. Pain mechanism is briefly introduced and a hierarchical stimulation model from coil stimulation to neuronal firing is proposed. Electromagnetic field distribution for circular, figure of 8 and Magnetic resonance coupling figure of 8 coils are analyzed with respect to the variation of stimulation parameters such as distance between coils, stimulation frequency, number of turns and radius of coils. MRC figure of 8 coils were responsible for inducing the maximum Electric field for same amount of driving current in coils. Variation of membrane potential, ion channel conductance and neuronal firing frequency in a pyramidal neuronal model due to magnetic and acoustic stimulation are studied. The frequency of neuronal firing for cortical neurons is higher during pain state, compared to no pain state. Lowest neuronal firing frequency 18 Hz was found for MRC figure of 8 coils, compared to 30 Hz for circular coils. Therefore, MRC figure of 8 coils are most effective for modulation of neuronal firing, thereby achieving pain relief in comparison to other coils considered in this study.*

## 1 Introduction

Pain could be termed as an unpleasant sensory and emotional experience which could involve actual or potential tissue damage, or be described in terms of such damage. Approximately 1.5 billion people worldwide experience chronic pain at some point in their lives. The American Pain Society stated the total annual incremental cost of health care due to pain ranges from $560 billion to $635 billion. There have been different kinds of medications which have used for the treatment of pain relief in the form of analgesics, invasive/non-invasive stimulation options and traditional Chinese medicine (Acupuncture). However, there still lies a pressing need for an effective and convenient system which could be used at the convenience of homes, reduce the costs associated with the treatment and lead to improvement in quality of life [1].

The most commonly accepted theory regarding the mechanism of pain is the gate theory of pain. The theory states that nerve impulses flow from peripheral nerves to CNS and a gate controls the flow of nerve impulses, "closed state" reducing the pain sensation. Nociceptors or pain receptors are free nerve endings, activated by biological, chemical, thermal, mechanical or chemical stimuli. Generally, two types of fibers, large A delta fibers (fast pain) and C fibers (slow or chronic pain) are used for transmission of pain impulses via the Dorsal Root Ganglion (DRG), which are responsible for relaying the nociceptive information to the spinal cord. The spinal cord then transmits the information to the thalamus, where it is perceived. The pyramidal neurons at Layer V in cortex get activated and show increased activity, synonymous to higher perception of pain. The ascending and descending pathway are responsible for the perception of pain from noxious stimuli and for pain modulation. Electrical stimulation of these neurons facilitate specific neurotransmitter release (endorphin, serotonin, GABA) at the neuronal membranes, which help in modification of action potential at the membranes. This modulates the neuronal firing activity of pyramidal neurons in cortex, where reduced neuronal activity signifies reduced perception of pain [2-4]. Fig. 1 briefly describes the pain mechanism for the reader to have a basic understanding.

In view of the large segment of population suffering from chronic pain, different kind of neural stimulators have been used for treatment of chronic pain. Invasive stimulation techniques such as deep brain stimulation (DBS) have been used quite effectively for the treatment of pain with high specificity [5]. However, it has been known to produce side effects when stimulation occurs at non-optimal sites during implantation. Non-invasive stimulation techniques such as transcutaneous electric nerve stimulation (TENS) and Transcranial Magnetic Stimulation (TMS) have also been found to be in wide usage. TENS has been known to produce a surface effect without deep penetration, where electrodes are placed on the skin with a small current passed through the electrodes for stimulation (monophasic signal), thereby not being very effective [6]. TMS involves usage of a very large current in the coils (~2-3 kA) due to usage of a biphasic signal, which leads to high power consumption. TMS coils are known to project a strong magnetic field at the cortex due to electrical currents which flow through the coils. These fields induce electric currents in cortical neurons, thereby hyperpolarizing axons which lead to modulation of neuronal firing, thereby achieving pain relief [7-8]. Pulsed RF signal used in this case for electrical stimulation has been previously found to be effective for pain relief applications [9]. Submillimeter coils or micro-coil arrays, with their small size have been recently developed to be implanted within the brain parenchyma to induce specific neuronal responses with change in spatial orientation of the coils [10]. Considering the disadvantages of the existing techniques, we focus on non-invasive electric stimulation using coils, where we try to achieve the same efficacy in terms of electric field induced, less power and higher safety due to reduced current driving the coils and a non-invasive approach.

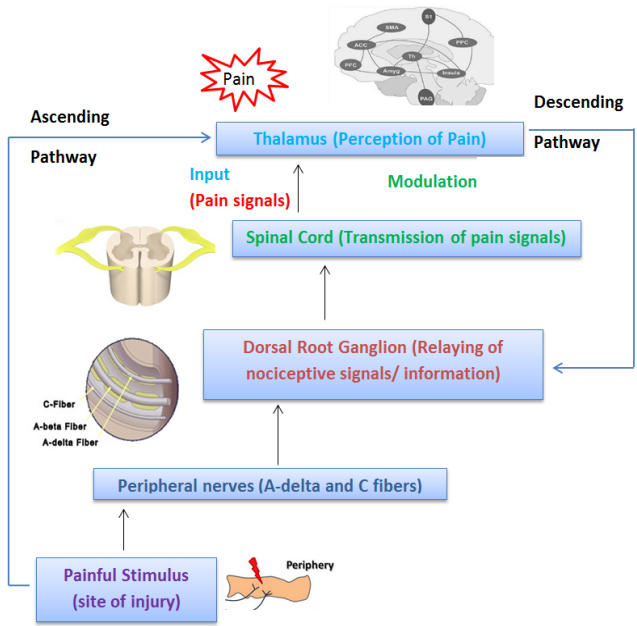

**Fig.1** Block Diagram for pain mechanism

This paper is aimed at evaluating the effects of electric field induced in pulsed RF magnetic stimulation for three different types of non-invasive coils: Circular, Figure of 8, and Magnetic Resonance Coupling (MRC) Figure of 8 coils with respect to change in various stimulation parameters such as size and shape of the coils, distance between the coils and stimulation frequency [11]. Besides, the effect of introducing non-linear acoustic propagation (ultrasound) in neurons and changes in action potential, conductance values of ion channels in a pyramidal neuron model due to stimulation from these coils are also studied. The paper will hypothesize the effect of modulation of neuronal firing due to both acoustic and electric stimulation, thereby achieving modulation of pain intensity.

## 2 Methods

### 2.1 Electromagnetic field distribution due to the coils

The magnetic field generated by these non-invasive coils, induces an electric field between the coils which is used for stimulation. Electric field E induced at the site of stimulation can be calculated by determining the derivative of the equation used for magnetic flux generated by the coil. This is given as:

$$E(r) = \frac{\mu_0 \omega N I}{4\pi} \int_C \frac{dl'}{|r - r'|}, \quad (1)$$

Where, $\omega$ is the angular frequency of the current, $I$ is the magnitude of current flowing through the coil, $N$ is the number of turns, $dl'$ is the differential coil element, $\mu_0$ is the permeability, $r$ and $r'$ are the position vectors of the observation point and that of the differential element $dl'$. Quasi-static conditions and negligible thickness are assumed while determining the electric field using the above formula [12, 13].

Three types of coils have been considered in this paper, which are circular coils, figure of 8 coils and magnetic resonance coupling (MRC) Figure of 8 coils for the analysis of Electromagnetic field distribution based on the variation of different stimulation parameters. The pictures for the various types of coils: circular, figure of 8 and MRC figure of 8 coils are shown in Fig 2(a), (b) and (c) respectively. Detailed construction of these coils, along with the directions for the axes representation in which the electric field is measured for the three types of coils has been discussed in detail in the earlier work [14]. The parameters studied here with respect to the coils for modulation of neuronal firing are distance between the coils (d), frequency of stimulation (f), number of turns (n) and radius of the coils (r). The corresponding change in electric field intensity with variation of radius of the coils from r=3 cm to r= 6cm, number of turns in the coils from 10 to 30, distance between the coils is varied from d=7 cm to d= 9 cm and the variation of frequency from 300 kHz to 600 kHz are analyzed. The coils are constructed with regards to achieving the desired electric field as that of deep brain stimulation, which has been used for chronic pain [15]. The parameters can be varied depending on the application being targeted. Fig. 3 shows the brief workflow of the coil set-up used to stimulate a neuron.

Besides, circuit simulation using LT Spice IV [Linear Technologies, 2015] was also performed by sweeping the frequency from 400 kHz to 500 kHz, where the peaks in current across the inductor coils is observed, thereby helping in determination of resonance frequency for the coils.

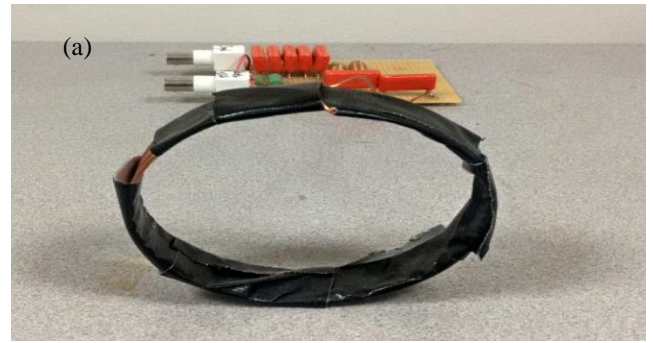

(a)

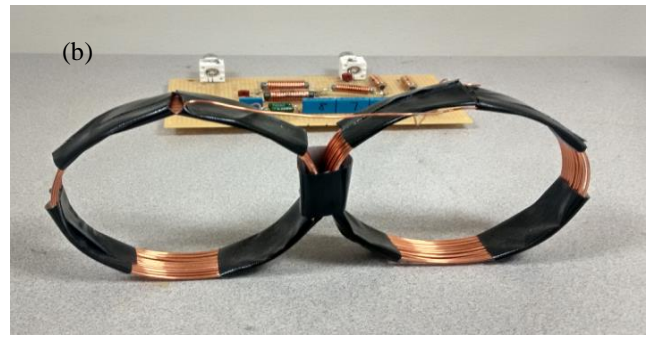

(b)

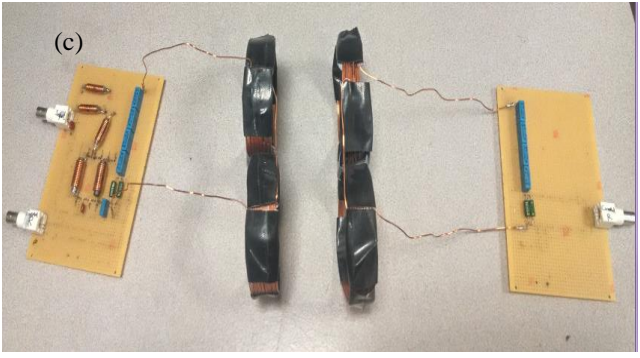

**Fig. 2** (a) Circular coil (b) Figure of 8 coils (c) MRC Figure of 8 coils

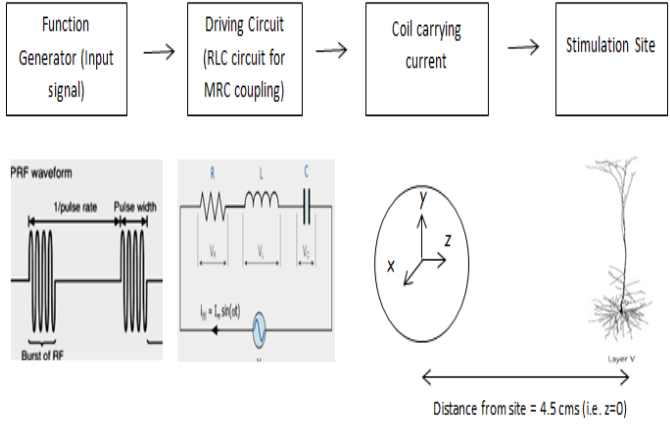

**Fig. 3** Workflow describing stimulating a cortical pyramidal neuron due to current carrying coils

*2.2 Acoustic propagation of an ultrasound wave*

In order to model the nonlinear propagation of a wave in the nearfield of an ultrasonic transducer it is necessary to allow for nonlinear propagation, diffraction, and attenuation (and dispersion). Different numerical approaches have been proposed depending on the nature of waveform and the geometry of the transducer. One of the most common approaches in this regard is the KZK [Khokhlov-Zabolotskaya-Kuznetsov] equation which is a nonlinear equation to account for the non-linearity and diffraction in sound beams [16]. Non-invasive brain stimulation using low frequency Ultrasound has already been reported to alter membrane conductance and modulate neuronal activity, which could potentially lead to pain relief [17].

Here, the neuronal stimulation due to an acoustic source is studied in terms of pressure peak amplitude, spatial and temporal waveforms and intensity, which could be used in conjunction with the magnetic stimulation from coils in the hierarchical neural stimulation model set-up (described in Fig 5) to enhance the stimulation effect in neurons.

If Z-axis is the direction of beam propagation and the transducer lies in the (x, y) plane normal to the Z-axis, the equation can be written as:

$$\frac{\partial^2 p}{\partial z \partial \tau} = \frac{c_0}{2} \nabla^2 + \frac{\delta}{2c_0^2} \frac{\partial^3 p}{\partial \tau^3} + \frac{\beta}{2\rho_0 c_0^3} \frac{\partial^2 p^2}{\partial \tau^2} \quad (2)$$

where, $p$ is the sound pressure, $c_0$ is the small signal sound, speed, $\delta$ is the sound diffusivity, $\beta$ is the nonlinearity coefficient, $\rho_0$ is the density and $\zeta$ is the retarded time and $\nabla_\perp^2$, is the Laplacian operator in the x-y plane. It is solved in the frequency domain using a finite difference scheme for forward propagation of the wave. The pressure wave is modelled as a Fourier series consisting of the fundamental and its harmonics. A set of equations for each harmonic at each grid point are written and solved by finite difference schemes [18, 19].

Integration of the axisymmetric KZK equation is performed in the frequency domain. As a result, spatial distribution of pressure of each harmonic is observed, with consideration of factors such as interference effects, power-law frequency-dependence of absorption beam diffraction, nonlinear effects of higher harmonic generation and the corresponding phase velocity dispersion. From these pressure fields the temporal average intensity is calculated. We model the system to our set-up where the US beam travels for a penetration depth of 8cm, i.e. the focal length of the transducer, where maximum vibration of the tissue occurs. The parameters used in simulation are highlighted in Table 1.

*2.3 Neuron Simulation for a cortical pyramidal neuron*

The neuronal simulation was performed on a layer V pyramidal neuron in cortex as shown in Fig 4. A pulsed RF electric field was applied to the neuron and the neuronal firing frequency was observed in the constructed model. The induced Electric field for the three different kinds of coils was calculated and the determined Electric field intensity was substituted in the Hodgkin Huxley equations shown below to determine the ionic conductance for the channels and the membrane voltage.

**Table 1** Parameters for an acoustic propagation in a neuronal tissue model

| Material | Parameter | Symbol | Value |
|---|---|---|---|
| 1-Water | Small Signal sound speed | $C_1$ | 1482 m/s |
| | Mass density | $\rho_1$ | 1800 kg/m$^3$ |
| | Absorption at 1 MHz | $\alpha_1$ | 0.217 dB/m |
| | Exponent of Absorption vs frequency curve | $\eta_1$ | 2 |
| | Non-linear parameter | $\beta_1$ | 3.5 |
| | Material Transition distance | $z$ | 5 cm |
| 2-Neuron | Small Signal sound speed | $C_2$ | 1482 m/s |
| | Mass density | $\rho_2$ | 1800 kg/m$^3$ |
| | Absorption at 1 MHz | $\alpha_2$ | 0.217 dB/m |
| | Exponent of Absorption vs frequency curve | $\eta_1$ | 2 |
| | Non-linear parameter | $\beta_1$ | 3.5 |
| Transducer | Outer radius | $a$ | 2.5 cm |
| | Inner Radius | $b$ | 1 cm |
| | Focusing depth | $c$ | 8 cm |
| | Frequency | $d$ | 1 Mhz |
| | Power | $P$ | 300 W |

$$C_m \frac{dV_m}{dt} = -g_l(V_m - E_l) - g_K(V_m - E_k)$$
$$- g_{Na}(V_m - E_{na}) + I_s \quad (3)$$

Where, $g_{Na}, g_K, g_l, E_{Na}, E_k, E_l$ are the sodium, potassium and leakage currents and the reversal potential of their corresponding magnitude of current. $C_m$ is the capacitance of the membrane, $I_s$ is the stimulation current and $V_m$ is the membrane voltage [20].

The nerve voltage membrane $V_m$ using the Voltage Kirchhoff Law for a simple biological membrane circuit can be determined by the relation:

$$\frac{V_m - V_{ind}}{R_i + R_e} + \frac{V_m C_m}{2} + \frac{V_m}{2R_m} = 0 \quad (4)$$

where, $R_m$ is the membrane resistance, $R_i$ and $R_e$ are the intracellular and extracellular resistance values of the membrane, $C_m$ is the capacitance value of the membrane and $V_{ind}$ is the induced voltage due to the magnetic stimulation circuit, which induces the circuit path linked by magnetic flux.

The above shown equation (3) was used to calculate the membrane potential. The membrane potential and firing frequency with respect to the induced Electric field intensity for the different coils was determined by implementing it in the NEURON 7.1 simulation platform [21]. The present study was done using cortical pyramidal neurons implemented using a current clamp approach to observe the neuronal firing frequencies and the membrane response. A current clamp at 100 Hz frequency was used, with differing values of magnitude of current clamp, based on the magnitude needed to induce the required Electric field. The simulation is performed for 500 ms with a time step of 0.025 ms. The neuronal firing in a pyramidal neuron model and the action potential peak across the neural membranes are studied for the three types of coils.

*2.4 Relationship between stimulation due to coils and the neuronal firing frequency: Hierarchical Model*

Various types of coils have been known to induce different magnitude and pattern of electric fields for magnetic stimulation depending on the shapes and sizes of these coils. As a result, these induced electric fields create varied effect on neurons, due to the magnitude of electrical field generated by these coils, with the same input parameters. Besides, the transducer helps in generating a strong acoustic force (maximum intensity) at the stimulation site, which aids in modulation of neuronal parameters.

It is well known that the neurons show increased neuronal activity in areas associated with pain such as Cortex [4]. Higher the neuronal firing frequency observed in cortex, more is the pain perceived by the human. This leads us to study the modulation of neuronal firing frequency due to the proposed stimulation approach.

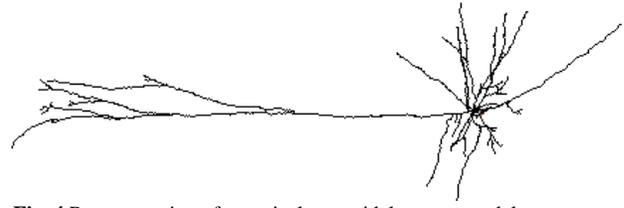

Fig. 4 Representation of a cortical pyramidal neuron model

| Top-bottom hierarchical neural stimulation analysis | | | |
|---|---|---|---|
| Coil Stimulation | Acoustic Stimulation | Biological Membrane | Neuronal Firing |
| Electric field induced and current peak values (circuit simulation) due to stimulation from circular, figure of 8 and MRC Figure of 8 coils | Acoustic non-linear propagation modelled using KZK [Khokhlov-Zabolotskaya-Kuznetsov] equation in neuron cells | Variation in Action Potential in terms of membrane voltage and conductance values across ion channels in neuron cells | Firing frequency in a pyramidal cortical neuron (Number of peaks in a specified time duration using current clamp) |
| Variation of E with change in parameters (Distance between the coils, Frequency, Number of turns and Radius of coils) | Variation of parameters like Axial/Radial pressure amplitude, axial peak intensity, temporal waveform peak observed with time | Modulation of the above parameters due to stimulation from all three coils, affecting the neuronal firing | Pain Intensity ∝ Firing Frequency (Modulation of firing frequency leads to reduced pain perception) |

Fig. 5 The model depicting the relation between stimulation due to the coils and the modulation of neuronal firing, leading to reduced pain

To do the same, we propose a hierarchical neural stimulation model. The electric field generated from coils, in combination with acoustic stimulation will modify the action potential and conductance values across ion channels.in neurons. This change will lead to modulation of neuronal firing frequency, where reduced firing frequency of these neurons will result in reduced perception of pain. This proposed hierarchical model has been shown in Fig 5.

### 3 Results

The relevant theory for determining the magnitude of Electromagnetic field intensity and the excitation of neurons due to electric stimulation from the three different types of coils has been described in the earlier section. The simulation results are presented in the forthcoming section.

*3.1 Electromagnetic field intensity distribution*

Firstly, the Electric field intensity induced for stimulation for the three sets of coils is shown in Fig 6. Here, x=0, y=0 and z=0 denotes a point which is at a distance of 4.5 cms away from the coils The detailed axes diagram and representation of the coils is shown in greater detail in [14]. The electric field is determined in each of the X, Y and Z directions by sweeping with the parameters as described earlier. The following are the parameters finalized for the representative application here to compare the fields due to these various coils as specified earlier: Distance between the coils = 9 cm, radius of coils = 4.5 cm, frequency of stimulation = 450 kHz and number of turns of coils = 20 [15].

Along the X and Y- axis, it is found that the Electric field decreases as we move away from the site of stimulation between the coils. Along the Z- axis, the electric field the coils, the electric field due to the MRC figure of 8 coils have the largest magnitude of Electric field as compared to the other coils, thereby making it more effective for neuronal modulation in terms of a deeper penetration effect as compared to the Figure of 8 coil and circular coil. For biomedical applications, it has been found that in case of the other coils, the magnitude of Electric field induced is way below the required target limit [150 V/m]. Also, significantly more amount of current needs to be passed in the driver for Figure of 8 coils (2 times) and circular coils (4 times), to obtain similar magnitude of electric field intensity as compared to MRC figure of 8 coils. This is due to the phenomenon of magnetic resonance coupling between the MRC Figure of 8 coils at the resonance frequency, where an equal amount of current is induced at the secondary coil, thereby doubling the electric field induced at the focal point, with the same driving current in the primary circuit as compared to other coils.

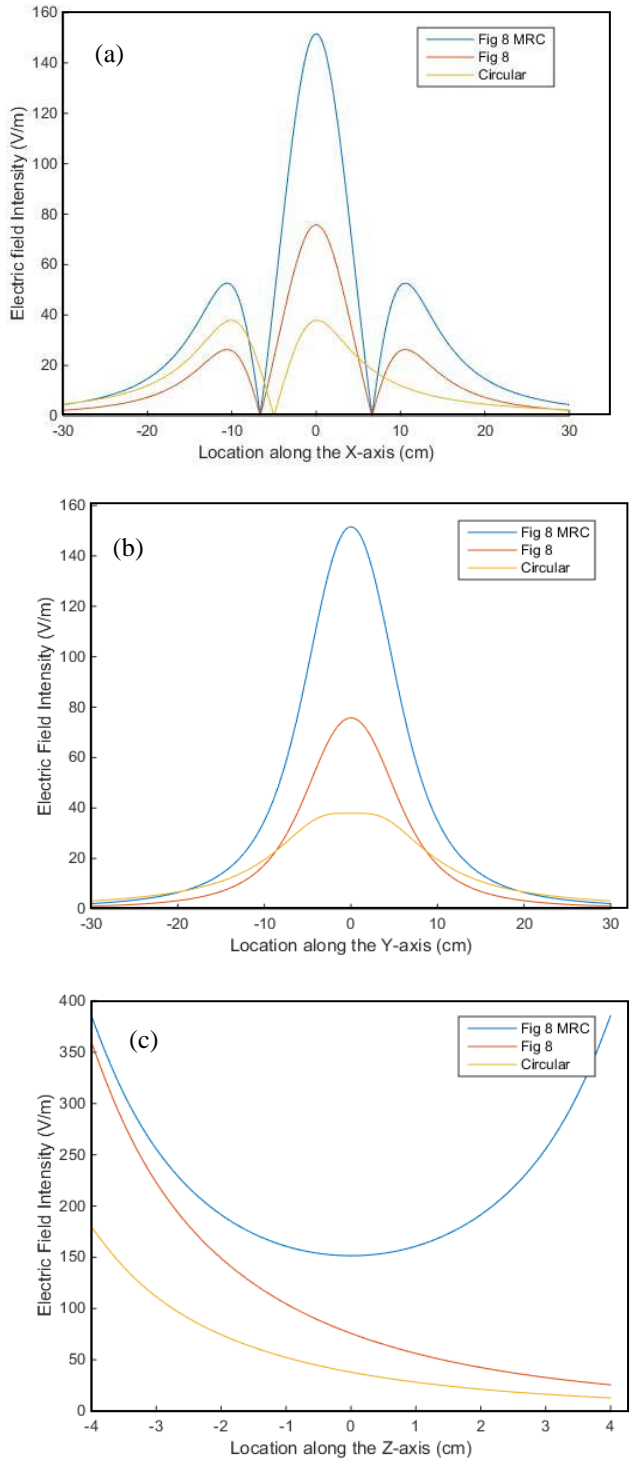

**Fig. 6** Electric field distribution due to the coils along the different axes (a) X-axis (b) Y-axis (c) Z-axis

*3.2 Comparison of Electric Field intensity with variation of stimulation parameters for all three types of coils*

The electric field induced for stimulation depends on various parameters as determined from the equation such as distance between the coils, frequency of stimulation, number of turns of the coils and radius of coils. In the section below, we analyze the change in Electric field intensity, with variation of the above mentioned parameters, when the other three factors are constant. The parameters are determined from the application to be focused on, based on the design of the stimulation system using these various coils. These calculations and results are obtained based on a unit magnitude of current flowing in the coils for representative purposes to understand the variation due to parameters. The current in the coils can be adjusted depending on the required Electric field which needs to be induced for the target application.

The changes in Electric field intensity with respect to variation of parameters of stimulation are as shown in Figures 7 (a)-(d). The results show a linear correlation, where the increase in parameters along the X-axis leads to an increase in the Electric field induced along the Y-axis in the figures obtained. The electric field intensity at the location between the center of coils decreases as the coils are moved further apart. However, in the case of circular coils and figure of 8 coils here, there is only one coil used for stimulation, and the electric field induced at a distance D/2 is computed. It was found that the MRC figure of 8 coils have the highest induced electric field compared to circular coils and figure of 8 coils in all the cases of variation of stimulation parameters due to the magnetic resonance coupling between the coils, when the same primary current is used for driving each of the three coils. The electric field intensity along the X, Y and Z axis for the parameters of stimulation are shown in Fig 8 (a) - (c). Summarizing, the electric field induced increases with increase in radius, number of turns and frequency of stimulation, but decreases when the distance between the coils is increased.

increases as we move towards the coils for MRC figure of 8 coils due to equal current induced in the secondary coil, whereas, for the other coils, the Electric field induced decreases as we move away from the coils. The Electric field is found to be maximum at the center of the coils (penetration depth D/2 = 4.5 cm) from the coils for the MRC coils.

Based on the plots observed for different coils in X, Y and Z directions, we find that for the given magnitude of current in

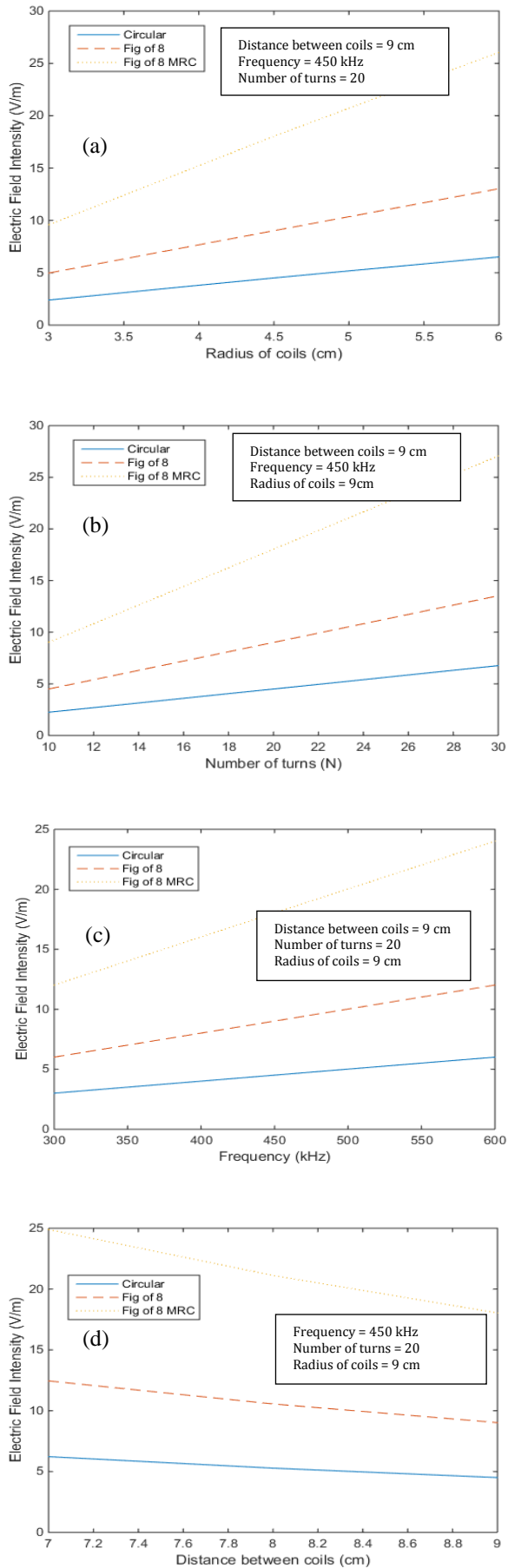

**Fig. 7** Variation of Electric field Intensity (a) change in radius of coils (distance between coils, frequency and number of turns are constants) (b) change in number of turns in the coils (distance between coils, frequency and radius of coil are constants) (c) change in frequency used for stimulation (distance between coils, number of turns and radius of coils are constants) (d) change in the distance between the coils (frequency, number of turns, radius of coils are constants)

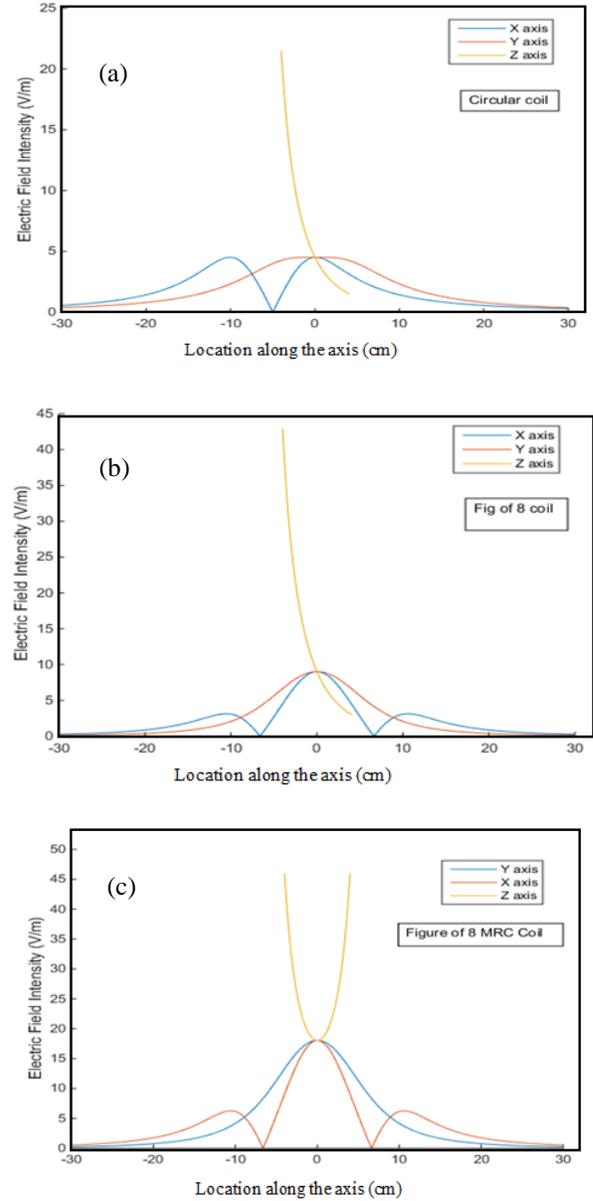

**Fig. 8** Plot of Electric field Intensity along the X, Y and Z axes for (a) circular coils (b) Figure of 8 coils (c) MRC Figure of 8 coils

*3.3 Electric Field Intensity for MRC Figure of 8 coils and variation with stimulation parameters*

MRC figure of 8 coils have been found to induce the maximum electric field intensity, compared to figure of 8 and circular coils for a similar magnitude of driving current in coils (when all other stimulation parameters are kept constant).

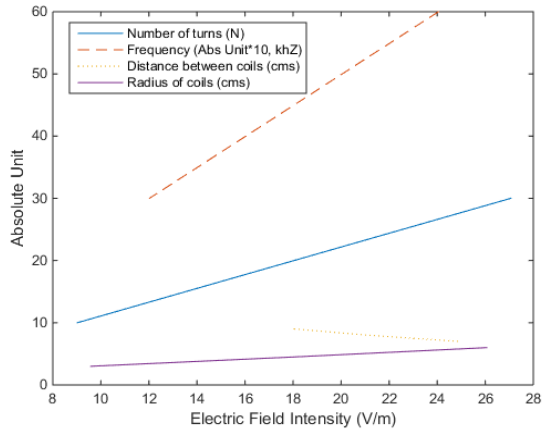

**Fig. 9** Variation of Electric field intensity with change in various stimulation Parameters for all the three coils

The slopes of the graphs describing the various stimulation parameters and their corresponding electric field intensity can be observed as shown in Fig 9. From the graph, the degree of variation of parameters for increasing/ decreasing the Electric field intensity to the required extent, for the modulation of neuronal firing can be determined. Optimization of the electric field to be induced can be performed by changing the stimulation parameters based on the requirements for the particular application. The source voltage can also be calculated and varied according to the requirements.

The variation of parameters along the X-axis show that with any change in parameter along the X-axis (such as between the coils, frequency, number of turns, distance or the radius of the coils) the difference in Electric field intensity (shown along the Y-axis) is more for MRC Figure of 8 coils as compared to the other coils, which shows that for small change in parameters, the MRC figure of 8 coils is more sensitive to enhancement of induced E field as compared to the other coils used for comparison in this study.

*3.4 Circuit Simulation results for coils with and without the use of Magnetic Resonance Coupling*

In this study, three types of coils have been considered which are MRC figure of 8 coils, Figure of 8 coils and Circular coils (as shown in Figure 2). The circuit simulation results obtained using LT spice where the frequency is swept across the range of 400 kHz to 500 kHz have been shown in Fig 10 (a) - (b). This circuit simulation results, is used to verify the concept of Magnetic Resonance Coupling for the MRC Figure of 8 coils, as compared to the other two coils, where only one coil is used. The latter set-up is similar to TMS, where a large current flows through one coil, for inducing the desired E field. In that case, only one peak will be observed, as compared to the MRC figure of 8 coils, where two peaks are observed due to the phenomenon of magnetic resonance coupling between the coils. This verifies well with the theory which states that due to magnetic resonance coupling, wireless power transfer occurs between the coils across the medium, which induces similar magnitude of current in the secondary coils, resulting in formation of two peaks across the two coils.

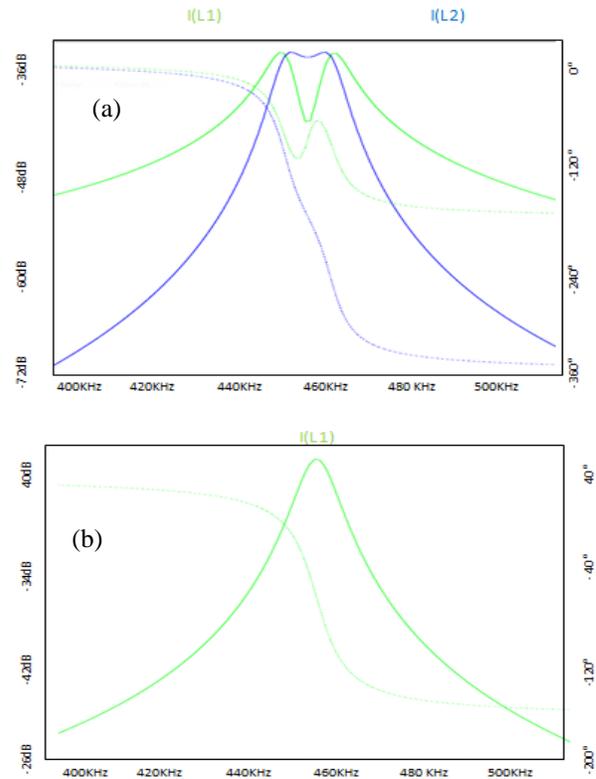

**Fig. 10** Circuit simulation results using LT Spice for (a) MRC Figure of 8 coils and (b) Figure of 8/Circular coils

*3.5 Acoustic characterization of focused Ultrasound fields for a cortical neuron Model*

In this case, stimulation using an ultrasound source has been modelled using KZK equation where the neuronal tissue response is predicted based on a similar set-up as compared to stimulation using the coils. The nonlinear propagation of an acoustic source is modelled in the nearfield of an ultrasonic transducer (where the neuron is present) to allow for nonlinear propagation, diffraction, and attenuation, resulting in modulation of neuronal firing, and changes in neuronal properties during the stimulation period. The results obtained below in Fig 11 (a) - (e) for the KZK equation show that axial pressure and intensity are maximum at the focus. Also, the peak positive and negative pressures tend to increase till the point it reaches the focus and then gradually reduces its value. The peak pressure in temporal waveform was found at the focal point.

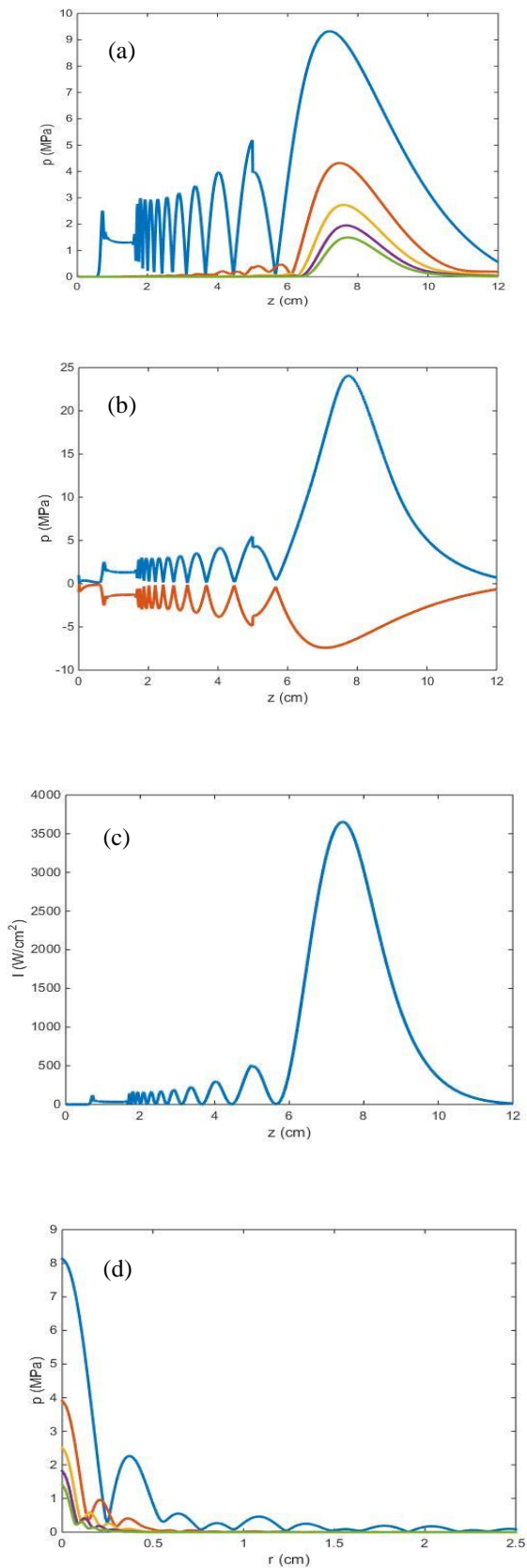

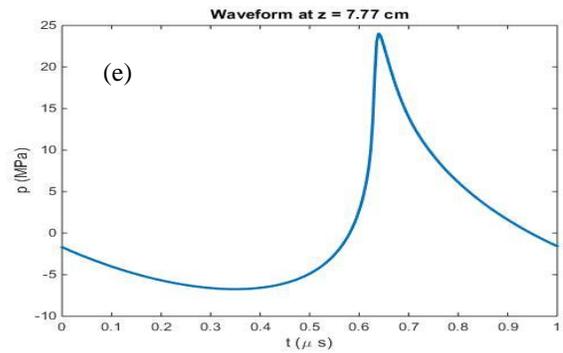

**Fig. 11** (a) Axial pressure amplitude of the first five harmonics (b) Axial peak positive and negative pressures (c) Axial intensity (d) Radial pressure amplitude of the first five harmonics at focus (e) Temporal waveform (on axis) at distance where peak pressure occurs.

The axial intensity and temporal waveform was found to be having the peak value at the focal point of the transducer, where maximum vibration occurs. Therefore, it can be hypothesized that the use of an ultrasound source in combination with the magnetic stimulation set-up of coils, can amplify the stimulation effect of the system, in terms of better efficacy of stimulation leading to greater neuronal modulation. This phenomenon is similar to the use of a low frequency ultrasound source for drug delivery or neuronal stimulation, where the vibration due to an acoustic source opens up the channels, where the drug is targeted to be delivered. Similarly, in this case the use of an ultrasound source based on the modelling parameters will enhance the modulation of the neuronal parameters at the membrane due to peak pressure, leading to reduced neuronal firing and reduced perception of pain signals.

*3.6 Action Potential and ion channel conductance variation in a pyramidal cortical neuron model due to stimulation from the three coils*

We know that with the variation of stimulation parameters for each of the coils, the electric field intensity induced for stimulation changes. This is most sensitive in case of MRC figure of 8 coils, where the change is more pronounced due to the structure of the MRC Figure of 8 coils. Here, we utilize the magnetic field from coils and acoustic field from transducer as discussed earlier, to create changes in membrane voltages and ion channel conductance values for a pyramidal cortical neuron model due to change in the induced E field. If we consider that all the coils are used for a similar application i.e. stimulating a specific neuron in Layer V of cortex, the magnitude of current induced due to the coils due to the three coils will differ, thereby leading to varied degree of modulation of ion channel conductance and action potential across neuronal membranes, as shown in the figures 12 and 13.

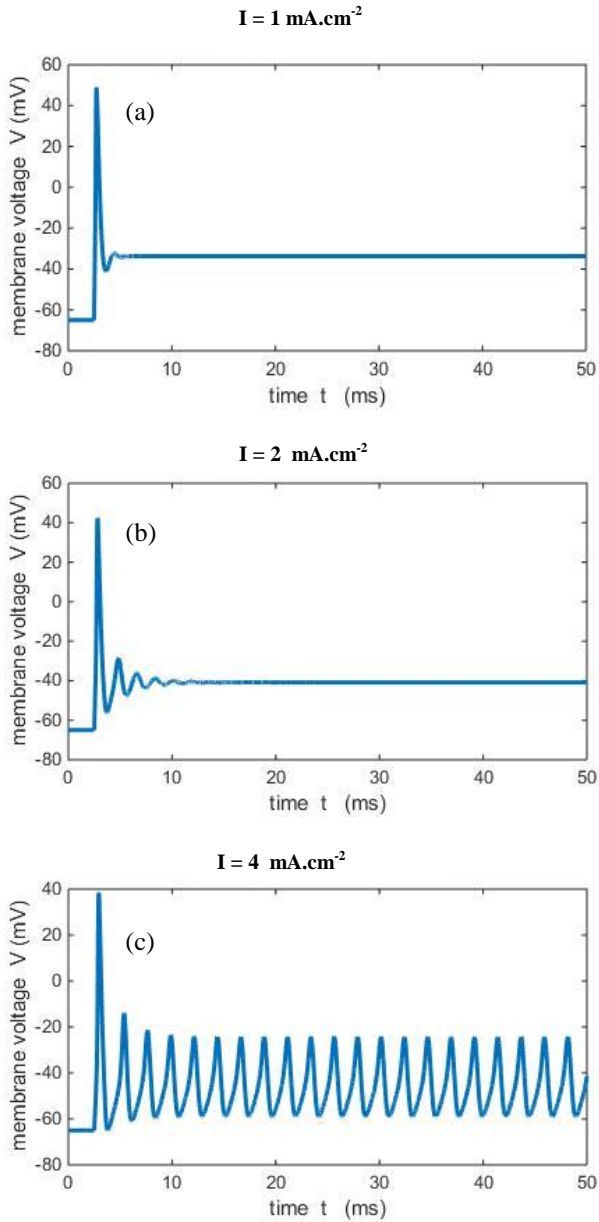

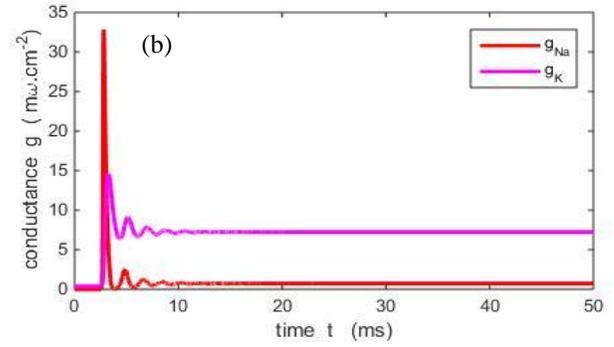

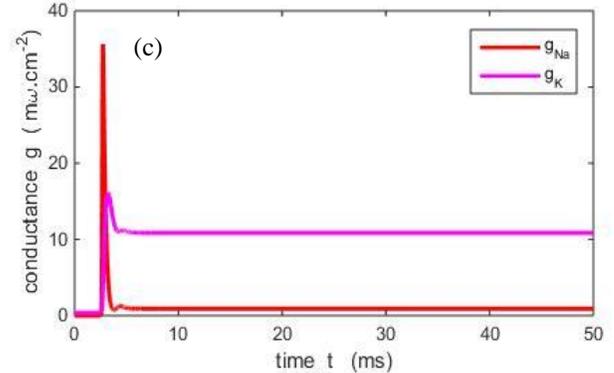

**Fig. 13** Conductance in ion channels due to stimulation from (a) Circular coils (b) Figure of 8 coils (c) MRC Figure of 8 coils

Figures 12 (a) - (c) show the variation of action potential with the change in induced current for stimulation. The figures denote that with increase in the current induced for fstimulation, the magnitude of action potential will decrease due to greater modulation of action potential across the neuronal membranes. Current induced is directly proportional to the electric field induced due to the coils, which approximately follows the ratio of 4:2:1 for MRC figure of 8 coils, figure of 8 coils and circular coils respectively. The corresponding variation of action potential and conductance values for ion channels in neuronal membranes are highlighted if a unit magnitude of current flows through the coils. The induced current can be changed depending on the application by varying the stimulation parameters discussed earlier. Fig 13 (a) - (c) show the variation in conductance values across the ion channels due to stimulation. Increased electrical field across the membrane and acoustic force leads to increase in influx of ions across the membrane, which is reflected in the increase conductance values of ion channels across the membrane.

*3.7 Neuronal firing frequency and action potential peak due to the stimulation from coils implemented in NEURON*

Figures 14 (a) - (c) show the neuronal firing in terms of a cortical pyramidal neuronal model implemented in NEURON 7.1 simulation platform. The neuronal firing frequency in terms of magnitude is found to be the maximum for circular coils, and the least for MRC figure of 8 coils. This implies that the maximum modulation of neuronal firing for the specified neuron model is achieved in case of the MRC figure of 8 coils, compared to the other coils. Figures 15 (a)- (c) show the corresponding action potential peak in soma due to the neuronal firing observed. It was found that the MRC figure of 8 coils had the least magnitude of peak action potential in the neuronal membrane, thereby establishing further that the maximum modulation of action potential i.e. pain perception is achieved due to MRC figure of 8 coils during stimulation, as compared to other coils.

**Fig. 12** Action potential variation for current induced due to (a) circular coil (b) Figure of 8 coils (c) MRC figure of 8 coils

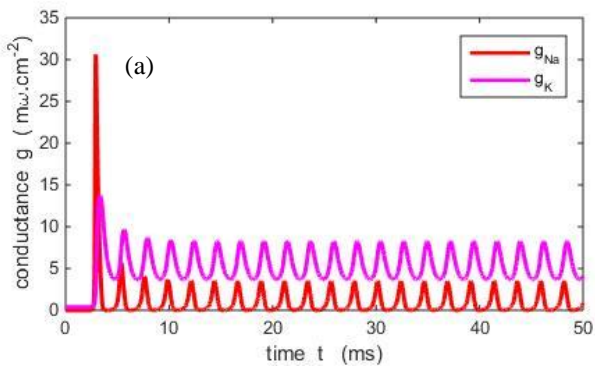

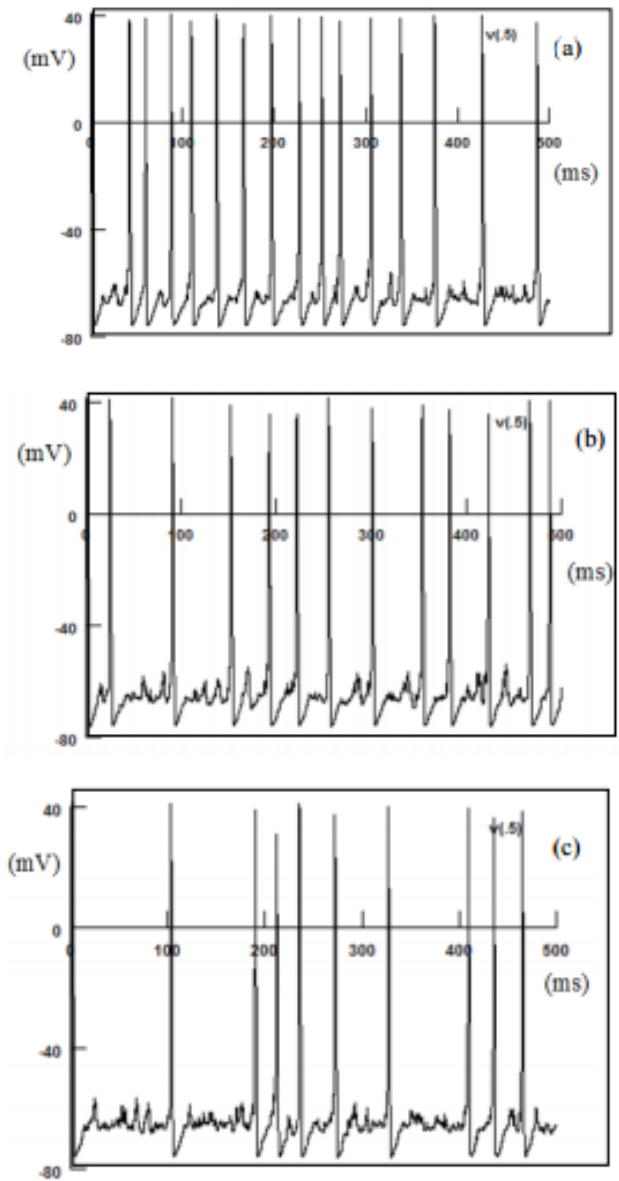

**Fig. 14** Frequency of firing of neurons in (a) Circular Coil: 30 Hz (b) Figure of 8 coils: 24 Hz (c) MRC figure of 8 coils: 18 Hz

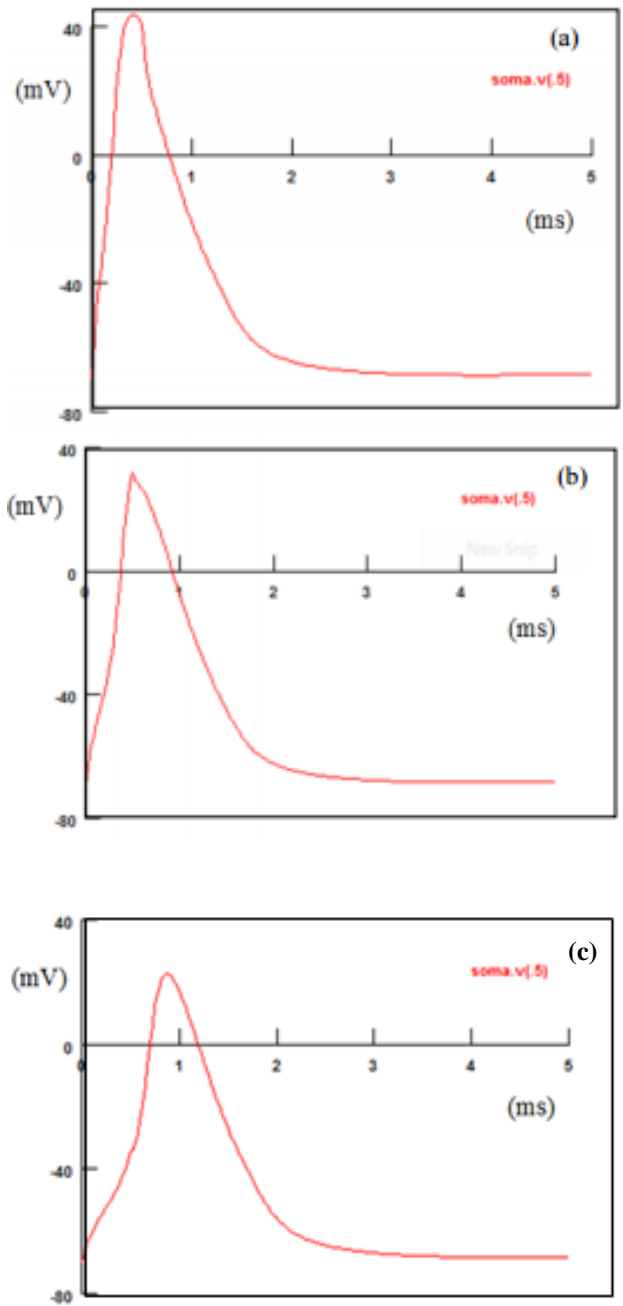

**Fig. 15** Action potential peak due to neuron firing in soma by stimulation using (a) Circular coils (b) Figure of 8 coils (c) MRC Figure of 8 coils

## 4 Discussions

This paper analyses four important findings in relation to the neural stimulation model proposed earlier: (a) Variation of coil design parameters and establishing efficacy in terms of magnitude of electric field induced for neuronal stimulation from circular, Figure of 8 coils and MRC figure of 8 coils; (b) intensity and pressure variation due to use of low frequency ultrasound for neuronal stimulation; (c) study of action potential variation and conductance values across neural membranes due to the hierarchical neural stimulation approach; and (d) measuring neuronal firing frequency, action potential modulation in a cortical pyramidal neuron model implemented in NEURON due to stimulation from these proposed coil designs, thereby determining the efficacy of stimulation for pain relief due to these coils. This study has contributed in observing the neuronal firing due to stimulation for a pyramidal cortical neuron. For magnetic nerve stimulation, a group of cortical neurons may respond differently depending on their threshold, however in general their firing frequency will be modulated, due to somatic depolarization, when it follows the proposed hierarchical neural stimulation model.

In the first part, the maximum electric field was found to be achieved in case of Magnetic Resonance Coupling (MRC) Figure of 8 coils due to magnetic resonance coupling between the coils at a specific frequency, increasing the electric field at the stimulation site, by using the same current used to drive the primary coil for all three coil set-ups. The distance between the two coils in the MRC coil design has been kept at 9cm, thereby enabling the maximum Electric field to be induced at a depth of 4.5 cm, which is the distance from the other two coils. This distance is significantly more than the penetration depth for the modulation of neuronal firing at the cortex at 3cm [15]. So, the current set-up of MRC figure of 8 coils can achieve much better stimulation effect in terms of neuronal modulation, being safer and using lesser current to drive the coils (low power) as compared to existing stimulators like TMS, which would use a high current (2 times magnitude) in the primary coil to achieve the same electric field. The comparison is performed in terms of the current needed for stimulation in both the butterfly coils and circular coils used with TMS. It was found that the current needed for stimulation in the MRC figure of 8 coils to induced an Electric field of 150 V/m is about 8-9A with a penetration depth of 4.5 cm, significantly lesser than the current needed in TMS coils(~2 kA). The voltage in case of TMS is in the range of kV, which is much higher, as compared to the case of MRC figure of 8 coils, where the source voltage is within the range of 100 V- 300 V depending on the excitation location of the coils, where $z$ varies from -4.5 cm to 4.5 cm, with $z = 0$, being the centre location at equal distances between the coils.

To check for the heating effect, we determine the specific absorption rate with a typical pulse with T= 100 us, Electric field induced = 150 V/m, cortex conductivity = 0.5 S/m, the maximum current density can be determined by:

$$J = \sigma E = 75 \ A/m^2$$

Then, the specific absorption rate (SAR) is given by [21]:

$$SAR = \frac{J^2 T f}{2 \sigma \rho}$$

Where the pulse frequency $f$= 100 Hz, tissue density $\rho$ = 1000 kg/m$^3$, SAR= 72.5 mW/kg, much below than the standard IEEE safety level, 400 mW/kg. This makes the proposed MRC Figure of 8 coils, much safer and effective compared to the traditional coils used in TMS, due to less power usage (smaller current in coils) and less heating effect with deeper penetration and increased specificity for neuronal modulation. In terms of coil stimulation effects, the following have been analyzed in the earlier sections: (1) Study of variation of stimulation parameters and their effect on induced electric field distribution for stimulation (2) Sensitivity study of the three coils for stimulation and (3) Comparison with TMS stimulation parameters, SAR for coils (to be within prescribed limit for safety issues).

The use of an ultrasonic transducer to achieve acoustic stimulation has also been explored. The parameters for stimulation using the KZK equation have been described to account for the non-linear propagation of US in neuronal tissue, where the axial peak intensity and pressure was found at the focal point of stimulation. This contributes in modulating the action potential across the neuronal membrane, where low frequency US have been found to be useful as highlighted in previous studies. The variation in action potential and ion channel conductance across neuronal membranes modelled using Hodgkin Huxley equations was studied. Due to the electric field induced from the coil which is used for stimulation, the action potential was found to reduce, and the ion channel conductance was found to increase in magnitude. The modulation of neuronal firing frequency was found to be maximum in case of MRC figure of 8 coils due to higher magnitude of electric field induced for neuronal stimulation. This takes place due to the stimulation current providing an influx of ions, which inhibit the action potential and increase the ionic conductance across the membrane.

This was further studied by implementation of a cortical neuron in NEURON, where the firing frequency of the neurons and action potential at soma was studied, in response to the electric field induced, after studying it at membrane level in the hierarchical model. The modulation of neuronal frequency and action potential across membranes (reduction in value) for the pyramidal cortical neuron due to stimulation from all the three coils are studied. It was found that the frequency of firing neuronal from an applied pulsed electric field reduces when MRC figure of 8 coils (18 Hz) are used as compared to a circular coil (30 Hz). This may be due to greater changes in conductance values for Na and K channels or the modulation of action potential for MRC Figure of 8 coils, due to the higher magnitude of electric field induced.

Since the neurons associated with higher pain sensation show increased activity in cortex, reduction of neuronal firing for cortical neurons would signify reduced perception of pain. The suppression of Action potential peak and modulation of neuronal firing is found to be the most when MRC Figure of 8 coils as compared to the other set of coils.

This comparison study for neuronal stimulation due to the coils and their efficacy for pain relief is performed by using the same set of initial conditions for each of the coils (power, input signal, circuit parameter, and same initial current driving the primary coil). Due to the structure and shape of the coils, each of them induced varied magnitude of electric fields, which is then used for stimulating a neuron and observing the neuronal firing frequency in a cortical pyramidal neuron implemented using NEURON. MRC figure of 8 coils are found to the most effective for modulating the firing

**Table 2** Comparison among the different coil designs where electric field was measured during experiment studies

| Type of Coil | Maximum |E| field at the penetration depth 4.5 cm | Frequency of firing of neurons in the cortical neuron mModel |
|---|---|---|
| MRC figure of 8 | 151.2 V/m | 18 Hz |
| Figure of 8 | 75.76 V/m | 24 Hz |
| Circular | 37.88 V/m | 30 Hz |

frequency, thereby being most suitable for pain relief applications.

## 5 Conclusions

In this paper, we propose non-invasive magnetic stimulation using coils and ultrasound transducer, and determine the efficacy of different types of coils such as circular coil, figure of 8 coils and MRC figure of 8 coils for stimulation in terms of modulation of neuronal firing. The change in electric field intensity with variation of parameters such as distance between the coils, frequency, number of turns and the radius of the coils is studied for all the three different types of coils. From the results, it was found that the MRC figure of 8 coils was found to be most effective in terms of the magnitude of electric field generation. The summary of comparison is presented in Table 2.

Besides this, acoustic propagation of an ultrasound transducer in a neuronal cell is also studied. Integration of the acoustic stimulation into the magnetic coil stimulation set-up may enhance the modulation of neuronal firing, i.e. lesser firing frequency due to E field induced from the coils, and the increased permeability of neuronal cells due to acoustic force. The variation of change in action potential and conductance values of ion channels due to stimulation from these coils is also studied. . Integration of the acoustic stimulation into the magnetic coil stimulation set-up may enhance the modulation of neuronal firing, i.e. lesser firing frequency due to E field induced from the coils, and the increased permeability of neuronal cells due to acoustic force.

Future work might involve performing patch clamp experiments using neurons to measure the neuronal response i.e. firing frequency and the conductance values of sodium and potassium channels due to stimulation from these sets of coils. Reduced neuronal firing may lead to lesser pain perception, which would be verified in-vitro. Animal experiments using mice can be further done to establish this fact, where their pain perception is measured using Von-Frey apparatus. Further experimental work can also be done with regards to combination of acoustic and magnetic stimulation, in a single modality in presence of an external magnet, to explore the use of magneto-acoustic signal for neuronal stimulation.

*Acknowledgements*- This research is supported by the Singapore National Research Foundation under Exploratory/ Developmental Grant (NMRC/EDG/1061/2012) and administered by the Singapore Ministry of Health's National Medical Research Council.